\documentclass[%
 reprint,
 superscriptaddress,
 amsmath,amssymb,
 aps,
]{revtex4-2}

\usepackage[utf8]{inputenc}

\usepackage{tikz}
\usepackage{adjustbox}
\usepackage[centering,hmargin=24mm,tmargin=25mm,bmargin=25mm]{geometry}
\usepackage{braket}
\usepackage{graphicx}
\usepackage{dcolumn}
\usepackage{bm}
\usepackage{color}
\usepackage{hyperref}
\usepackage{svg}
\hypersetup{colorlinks, 
    linkcolor={blue!75!black!80!yellow},
    citecolor={blue!75!black!80!yellow}, 
    urlcolor={blue!75!black!80!yellow}
    }

\usepackage{csquotes}
\MakeOuterQuote{"}
\makeatletter
\renewcommand\@make@capt@title[2]{%
    \@ifx@empty\float@link{\@firstofone}{\expandafter\href\expandafter{\float@link}}%
    \sffamily{\textbf{#1}}\@caption@fignum@sep#2
}%

\makeatother

\begin{document}

\title{Heterogeneous Multipartite Entanglement Purification for\\ Size-Constrained Quantum Devices}

\author{Stefan Krastanov}
\email{stefankr@mit.edu}
\affiliation{John A. Paulson School of Engineering and Applied Sciences, Harvard
University, Cambridge, MA 02138, USA}
\affiliation{Department of Electrical Engineering and Computer Science, Massachusetts Institute of Technology, Cambridge, MA 02139, USA}
\author{Alexander Sanchez de la Cerda}
\affiliation{John A. Paulson School of Engineering and Applied Sciences, Harvard
University, Cambridge, MA 02138, USA}
\author{Prineha Narang}
\email{prineha@seas.harvard.edu}
\affiliation{John A. Paulson School of Engineering and Applied Sciences, Harvard
University, Cambridge, MA 02138, USA}

\date{\today}

\begin{abstract}
The entanglement resource required for quantum information processing comes in a variety of forms, from Bell states to multipartite GHZ states or cluster states. Purifying these resources after their imperfect generation is an indispensable step towards using them in quantum architectures. While this challenge, both in the case of Bell pairs and more general multipartite entangled states, is mostly overcome in the presence of perfect local quantum hardware with unconstrained qubit register sizes, devising optimal purification strategies for finite-size realistic noisy hardware has remained elusive. Here we depart from the typical purification paradigm for multipartite states explored in the last twenty years. We present cases where the hardware limitations are taken into account, and, surprisingly, find that smaller `sacrificial' states, like Bell pairs, can be more useful in the purification of multipartite states than additional copies of these same states. This drastically simplifies the requirements and presents a fundamentally new pathway to leverage near-term networked quantum hardware.
\end{abstract}

\maketitle


Entanglement is one of the main non-classical resources that realize an advantage in quantum information processing. Bipartite entanglement, in particular, has inspired new communication protocols~\citep{ekert1991quantum,karlsson1999quantum}, quantum teleportation mechanisms~\citep{bennett1993teleporting,gottesman1999quantum}, quantum foundations experiments~\citep{bell1964einstein,aspect1981experimental,hensen2015loophole,giustina2015significant,shalm2015strong}, and has been the basis on which blind quantum computing, one-way computing, and cluster-state computing~\citep{raussendorf2001one} are built. The latter requires multipartite entangled stabilizer states, such as cluster states and GHZ states. The versatility of the entanglement resource makes its generation an important step in any quantum information processing architecture. However, generating Bell states or multi-qubit entangled states is also the lowest fidelity quantum operation in the current technology stack. While local gate fidelities are reaching levels of performance higher than 99\%, the network noise inherent to the distribution of entanglement can cause state fidelities lower than 90\%. To overcome this challenge, purification protocols have been proposed that sacrifice part of the noisy entanglement resources in order to detect errors and purify the leftover resource~\citep{dur2007entanglement,aschauer2005quantum,deutsch1996quantum,bennett1996purification,bennett1996mixed,dur1999quantum,dur2007entanglement,fujii2009entanglement,nickerson2013topological,nickerson2014freely,nigmatullin2016minimally,krastanov2019optimized,zwerger2018long,murao1998multiparticle,maneva2002improved,aschauer2005multiparticle,dur2003multiparticle,goyal2006purification,kruszynska2006entanglement,wallnofer2019multipartite,de2020protocols,zhou2016purification,zhou2016purification,zhou2020purification}.

\section{Introduction}

A typical premise in these purification methods is that the entangled state needs to fulfill a set of parity constraints (i.e., the entangled state is an eigenstate of a set of stabilizer operators) which have to be verified as part of the purification procedure. The same premise underlies classical and quantum error correction, which leads to interplay between the two topics~\citep{aschauer2005quantum,dur2007entanglement}. However, in the case of entanglement purification, we usually limit ourselves to only attempting to detect errors and discard faulty states, not correct them. The main hurdle in verifying such constraints in an entangled state is that they require non-local operations, as they span qubits spread among distant nodes. The standard solution is the use of circuits in which a second possibly-noisy entangled state is sacrificed in order to perform the non-local parity measurement. This can be seen as a continuation of other quantum-teleportation-based protocols, where a non-local resource state is expended in order to perform a non-local operation. As long as the fidelities of the input states are above a certain threshold, the information gained from the non-local measurement can increase the certainty that the purified state is error-free, conditional on the measurement reporting the expected parity.

An ideal purification protocol would sacrifice the smallest number of entangled states for the greatest increase in fidelity among the purified states. In the case of bipartite entanglement (i.e., Bell pairs), as long as the local nodes perform perfect noise-less operations and are not constrained in the number of entangled pairs they can share, a ``perfect'' solution is available. Namely, the hashing method~\citep{bennett1996mixed} provides the most efficient way in which one can extract information about the errors that have occurred on the states to be purified. The question becomes substantially more complex if we consider realistic local hardware. Given the diversity of possible hardware imperfections, there is no single optimal protocol. On one hand, the gate imperfections impose an upper bound on the fidelity of the purified state, unless scalable error correction and quantum repeaters with encoding are employed~\citep{jiang2009quantum}, which obviates the need for purification. On the other hand, the finite size of a quantum register limits the depth and width of the purification circuit. These trade offs make the otherwise perfect hashing method very unfavorable if not impossible. A family of hand-optimized protocols have been proposed since then~\citep{nickerson2013topological,nigmatullin2016minimally}, and general circuit optimization methods have been suggested~\citep{krastanov2019optimized,jansen2021enumerating}, designing circuits with awareness of the various noise processes, thus providing a practical solution to this problem.

Similarly, in the case of multipartite entanglement purification~\citep{murao1998multiparticle,maneva2002improved,aschauer2005multiparticle,dur2003multiparticle,goyal2006purification,kruszynska2006entanglement,wallnofer2019multipartite,de2020protocols,zhou2017polarization}, a variant of the hashing method provides the most efficient possible way to extract information about errors that might have been inflicted upon the states to be purified~\citep{bennett1996mixed,aschauer2005multiparticle,kruszynska2006entanglement}. However, this generalization of the hashing method requires not only perfect gates and quantum registers of unbounded size, but also a very particular set of sacrificial entanglement resources. While it is probable that once we have scalable error-corrected quantum hardware the hashing method will be used at the upper logical levels, hashing is much too demanding to be practical at the lower levels running on noisy small quantum nodes.

In this letter we showcase a number of small alternative circuits with much lower resource demands. In particular, our designs require simpler sacrificial entangled states than what is required by the typical multipartite entanglement purification protocols. When taking into account the limitations of near-term quantum hardware, these alternatives become much more practical than established approaches. We also exploit possible asymmetries in the way multipartite entanglement is generated and distributed over a network. This letter is organized as follows. We start with a brief overview of the mathematical language used to describe purification circuits - in particular, the stabilizer formalism~\citep{gottesman1998heisenberg}. Next, we discuss how raw multipartite entanglement is distributed in a network and the established methods for its purification. Following that, we introduce and evaluate our small heterogeneous purification circuits. We conclude by discussing the usability of our approach when applied to near-term quantum devices.

\begin{figure}
    \centering
    \includegraphics[scale=0.45]{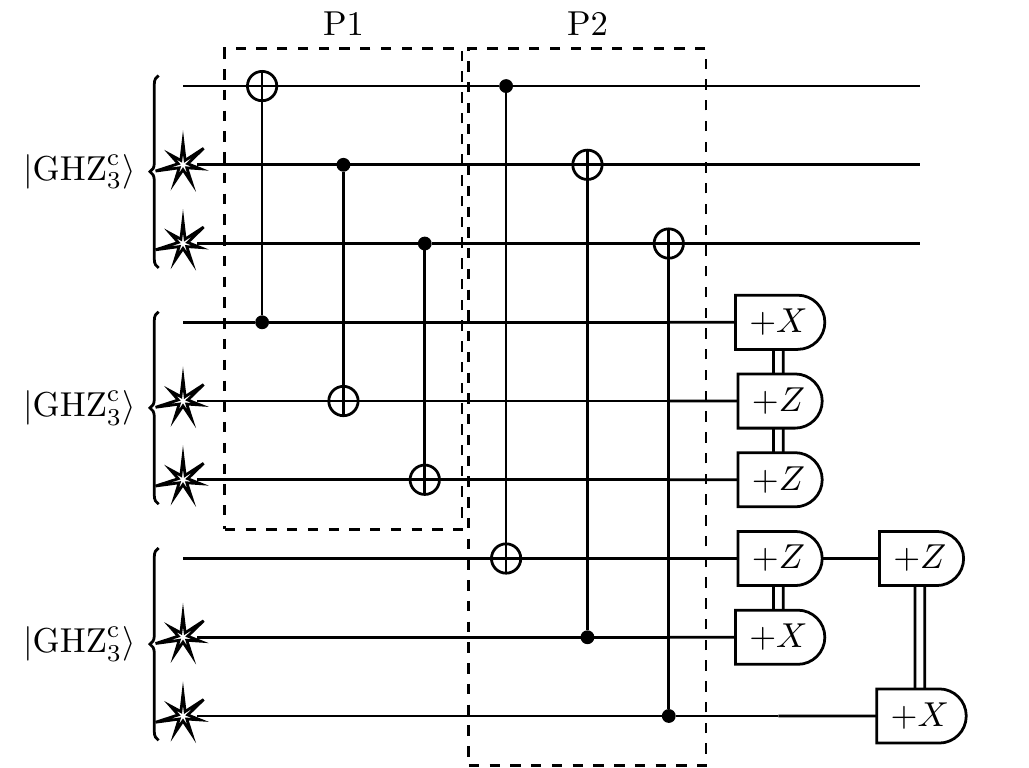}
    \includegraphics[scale=0.45]{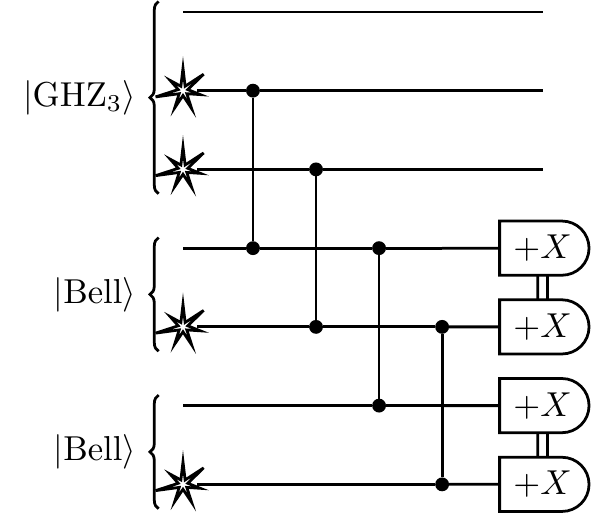}
    \caption{Purification circuits for 3 qubit GHZ states. On the left is the standard two-stage circuit out of which the hashing method can be built. It sacrifices an entire GHZ state at each stage. We denote the stages as P1 and P2 (a full implementation of the hashing method would sacrifice an asymptotic number of GHZ states in repeated applications of P1 and P2). On the right is our much simpler circuit sacrificing only two Bell pairs for the same final fidelity and higher success probability (or yield). Our circuit not only requires much simpler resources (4 qubits in two Bell pairs instead of 6 qubits in two GHZ states) but, as it is shorter, it is less susceptible to gate errors. The vertical classical communication (double lines) between the measurement symbols denotes the sets of classical bits whose parity needs to be checked in order to detect errors. Stars denote the qubits that need to be transmitted over the network, thus being subjected to noise. Measuring stabilizers that involve mostly the noisier qubits transmitted over the network provides for higher error detection efficiency.}
    \label{fig:3qs}
\end{figure}

\begin{figure*}
    \centering
    \includegraphics[width=0.31\textwidth]{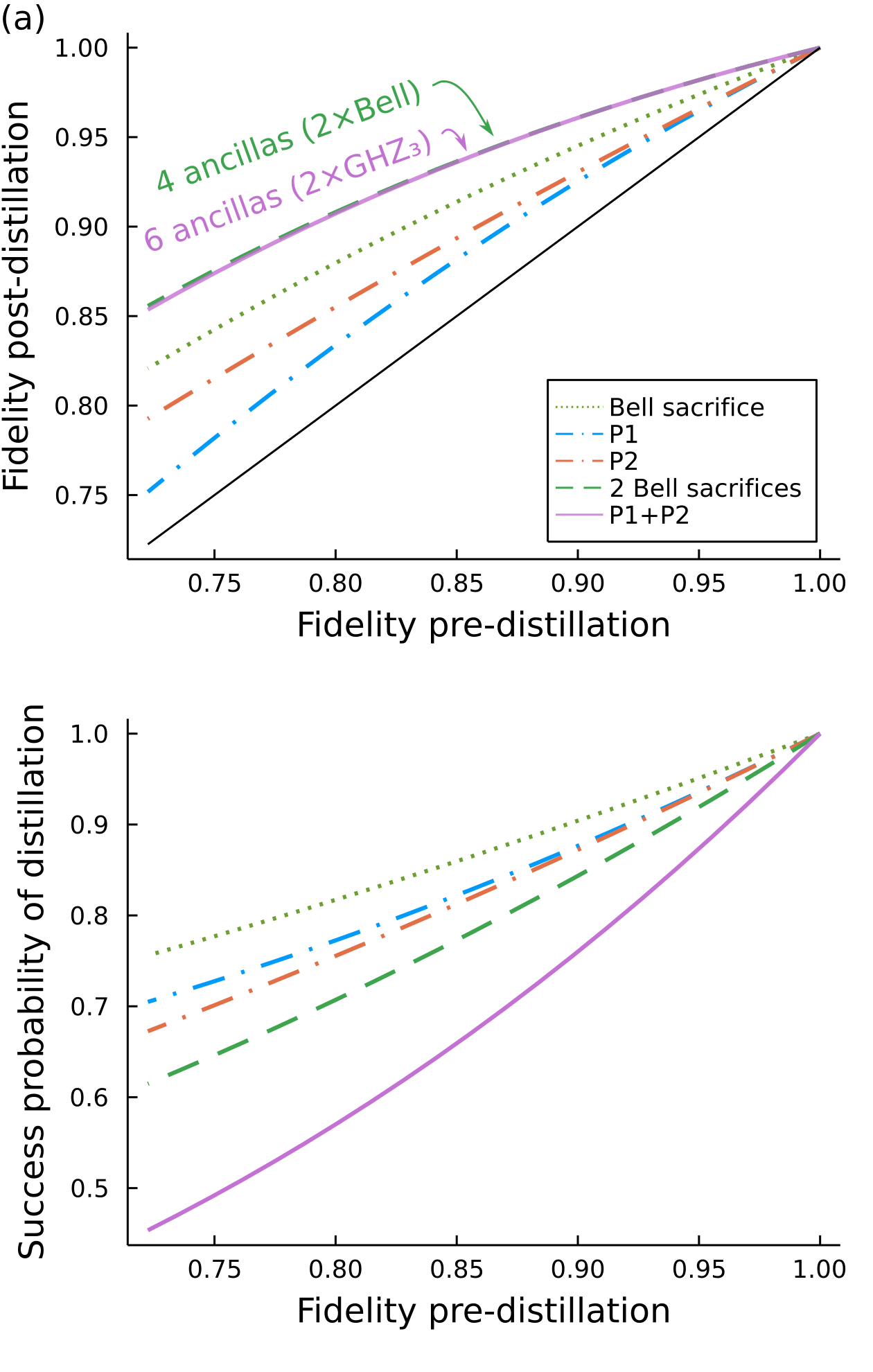}
    \includegraphics[width=0.31\textwidth]{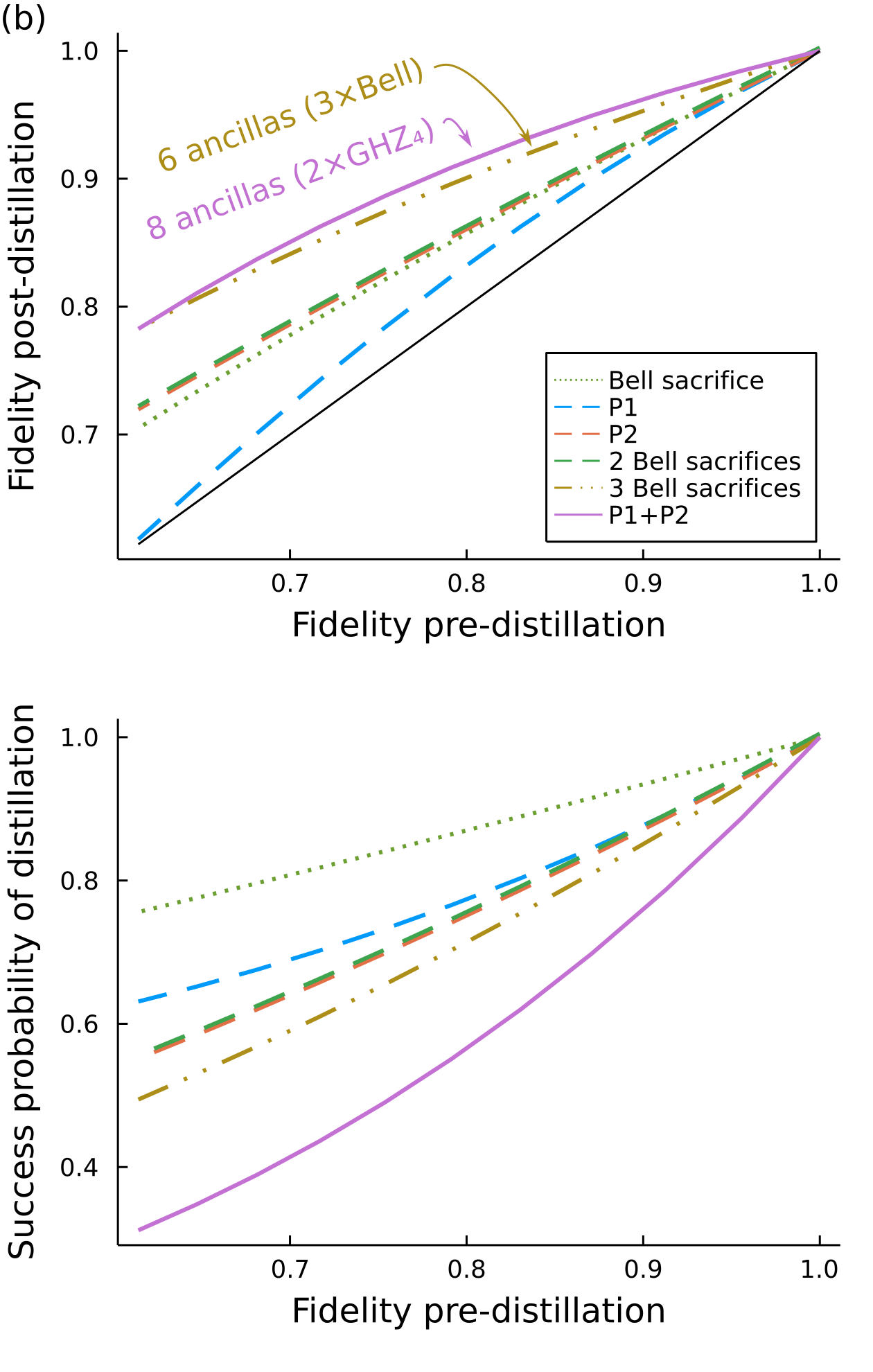}
    \includegraphics[width=0.31\textwidth]{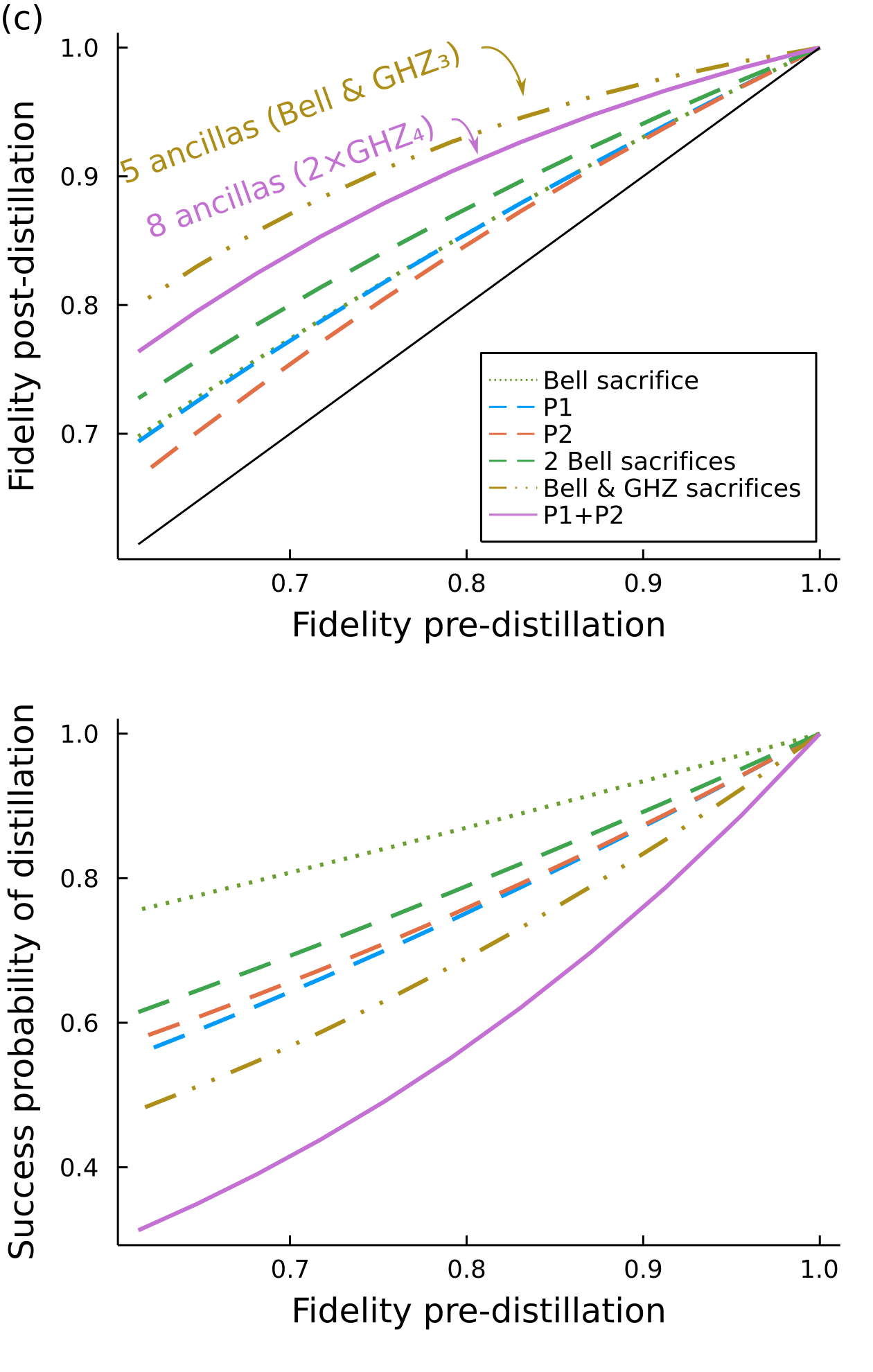}
    \caption{Performance comparison of various purification circuits. From left to right, the columns represent purification of (a) a single 3-qubit GHZ state, (b) a 4-qubit GHZ state, and (c) a 4-qubit linear cluster state. The top figures give the fidelity of the purified state, while the bottom figures give the probability of success for the purification circuit. In purple (P1+P2) is the typical "one block of the hashing method" considered in the literature, which we compare against. As the circuits are heterogeneous, one can not provide a well-defined yield. Rather, the main observation is that for short circuits performing the sacrifice of a small number of Bell pairs can provide higher final fidelity and a better probability of success than performing the sacrifice of the same number of large, expensive multipartite entangled states. The horizontal axis corresponds to the initial fidelity of the state to be purified. All circuits assume all entangled resources are generated locally on one of the nodes and then distributed over the network. All qubits sent over the network suffer the same rate of depolarization noise. The P1 and P2 purification circuits are the two stages of purification used in the standard hashing method. "P1+P2" denotes both stages being performed (and two multipartite states being sacrificed). In the case of purifying 4-qubit states we have included curves for the best circuits using 2 Bell sacrifices, but we do not discuss them in the main text as they do not show exceptional performance (they are available in the documentation of QuantumClifford.jl).}
    \label{fig:perf}
\end{figure*}

\section{Purification Circuits}

\subsection{A General Description}

Purification schemes are typically devised for one or more copies of a desired stabilizer state $\ket{\mathrm{P}}$ to be purified by sacrificing a number of other stabilizer states $\{\ket{\mathrm{S}_1},\ket{\mathrm{S}_2},\dots\}$. As the qubits of each of these states are spread among distant nodes, only local operations can be performed between them. The particular local operations that enable the purification, $\hat{U}$, are idempotent if no errors were present: $\hat{U} (\ket{\mathrm{P}}\otimes\ket{\mathrm{S}}) = \ket{\mathrm{P}}\otimes\ket{\mathrm{S}}$. To facilitate error detection, that same operation is meant to "infect" the sacrificial state $\ket{\mathrm{S}}$ if the state $\ket{\mathrm{P}}$ has suffered a given error $\hat{E}$ (e.g., a bit flip on one of the qubits): $\hat{U} (\hat{E}\ket{\mathrm{P}}\otimes\ket{\mathrm{S}}) = \hat{E}_{\mathrm{p}}\ket{\mathrm{P}}\otimes\hat{E}_{\mathrm{s}}\ket{\mathrm{S}}$. By performing local measurements on the state $\hat{E}_{\mathrm{s}}\ket{\mathrm{S}}$ and observing that the correlations between them are inconsistent with the state being $\ket{\mathrm{S}}$, we know an error is detected, and the purified-to-be state has to be discarded.

Designing a good purification scheme requires finding the locally constrained circuit $\hat{U}$ and the set of correlation measurements which provide the greatest chance for detecting errors on $\ket{\mathrm{P}}$, without letting errors on $\ket{\mathrm{S}}$ backpropagate to it. As we will see, this includes the judicious choice of sacrificial states whose preparation does not contribute to the infidelity and overhead of the purification procedure. One of the most common approaches involves using gate teleportation techniques, where the sacrificial state is effectively used to perform a non-local stabilizer measurement on the purified-to-be state. This would require the use of Bell states~\citep{de2020protocols} or, more generally, GHZ states as the sacrificial resource. This type of non-local stabilizer measurement is conceptually very similar the local stabilizer measurement in quantum error correction. In the case of local measurements, a single ancillary qubit is prepared in the $\mid0\rangle+\mid1\rangle$ state, followed by two-qubit operations with the qubits whose stabilizer is to be measured; these two-qubit operations accumulate phase on the ancillary qubit; depending on whether the parity of the stabilizer measurement is trivial or not, that phase will be $0$ or $\pi$, providing for a non-demolition parity measurement. Early on, a more fault-tolerant method was suggested~\citep{shor1996fault}, in which instead of a single ancillary qubit, multiple ancillary qubits forming a GHZ state are used. Besides being fault-tolerant, this method (when coupled with classical communication) provides the non-local parity measurement under consideration here.

The more powerful hashing method uses bilateral XOR gates between copies of the same entangled state (or a similar state in the case of general graph states~\citep{kruszynska2006entanglement}) in order to efficiently extract information about the probability distribution of errors in the set of purified states. However, the constraints imposed by both of these approaches are inappropriate when devices of finite size are considered, especially when infidelities in the local operations can not be neglected. For such noisy finite circuits a less constrained circuit $\hat{U}$ can perform significantly better at much lower resource requirements as seen in multiple works on optimal bipartite entanglement purification~\citep{nickerson2013topological,nigmatullin2016minimally,krastanov2019optimized}. Similarly, the multipartite purification circuits we propose have significantly lower resource overhead compared to the known asymptotic method when constrained to small near-term hardware.

\subsection{The Preparation of Sacrificial Entangled Resources}

It is important to discuss the way the initial multipartite entangled states are generated before optimizing a circuit for their purification. Bipartite entangled pairs can be generated either by a heralding operation where light interacts with both nodes and the path information is erased before observing the light, or by preparing the Bell pair locally on one of the nodes and sending one of the qubits over the network to the other node. We will focus on correcting the errors caused by the network transfer, affecting only the qubits actually transmitted over the network. Multipartite entangled states can also be created through heralding~\citep{walther2007heralded} or by having them locally generated and then having all but one of their qubits distributed over the network. Lastly, multipartite entanglement purification can also be done by first purifying Bell pairs and then fusing them into the larger multipartite states we desire, but it is established that that method does not perform as well as the direct multipartite purification, e.g. the hashing method~\citep{aschauer2005multiparticle}. In the case of GHZ states, purifying them directly by the use of Bell states was recently studied~\citep{de2020protocols}.

Crucially, the purification circuits we propose will be evaluated for the circumstances where the entangled states are generated locally and then all but one of the qubits involved are sent over the network to other nodes, as done in recent microwave qubit experiments~\citep{zhong2020deterministic}.  All but one of the qubits of each entangled state suffer errors. Exploiting this asymmetry is a significant factor in the effectiveness of our circuits. As an example, consider comparing a 4-qubit sacrificial GHZ resource to two sacrificial Bell pairs. The number of qubits is the same, and the Bell pairs "contain" less entanglement in absolute terms, but only two of the Bell pair qubits would have to had been sent over the network, while three of the GHZ qubits have had to travel to other nodes. For all examples in this letter, the noise experienced by qubits sent over the network is unbiased depolarization.

\begin{figure*}
    \centering
    \includegraphics[scale=0.55]{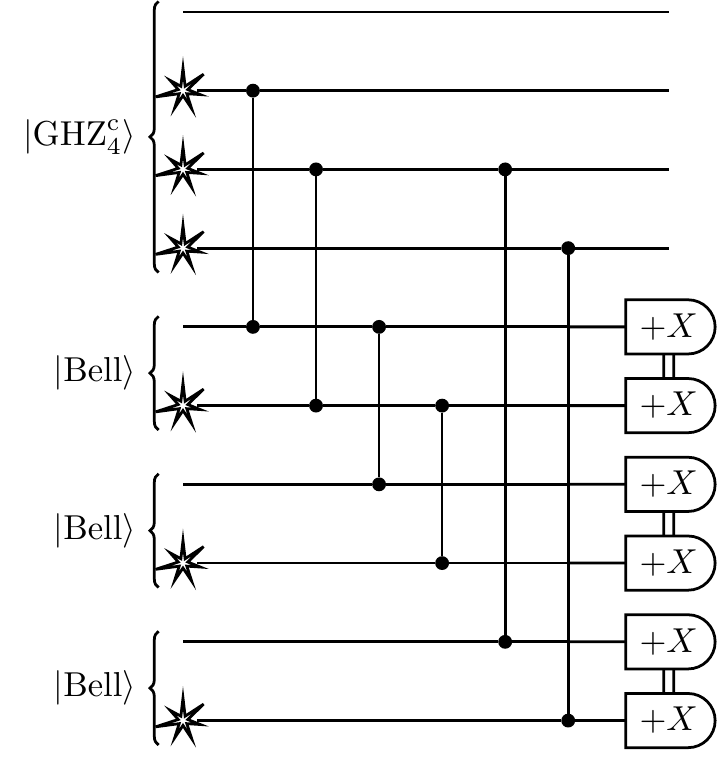}
    \includegraphics[scale=0.55]{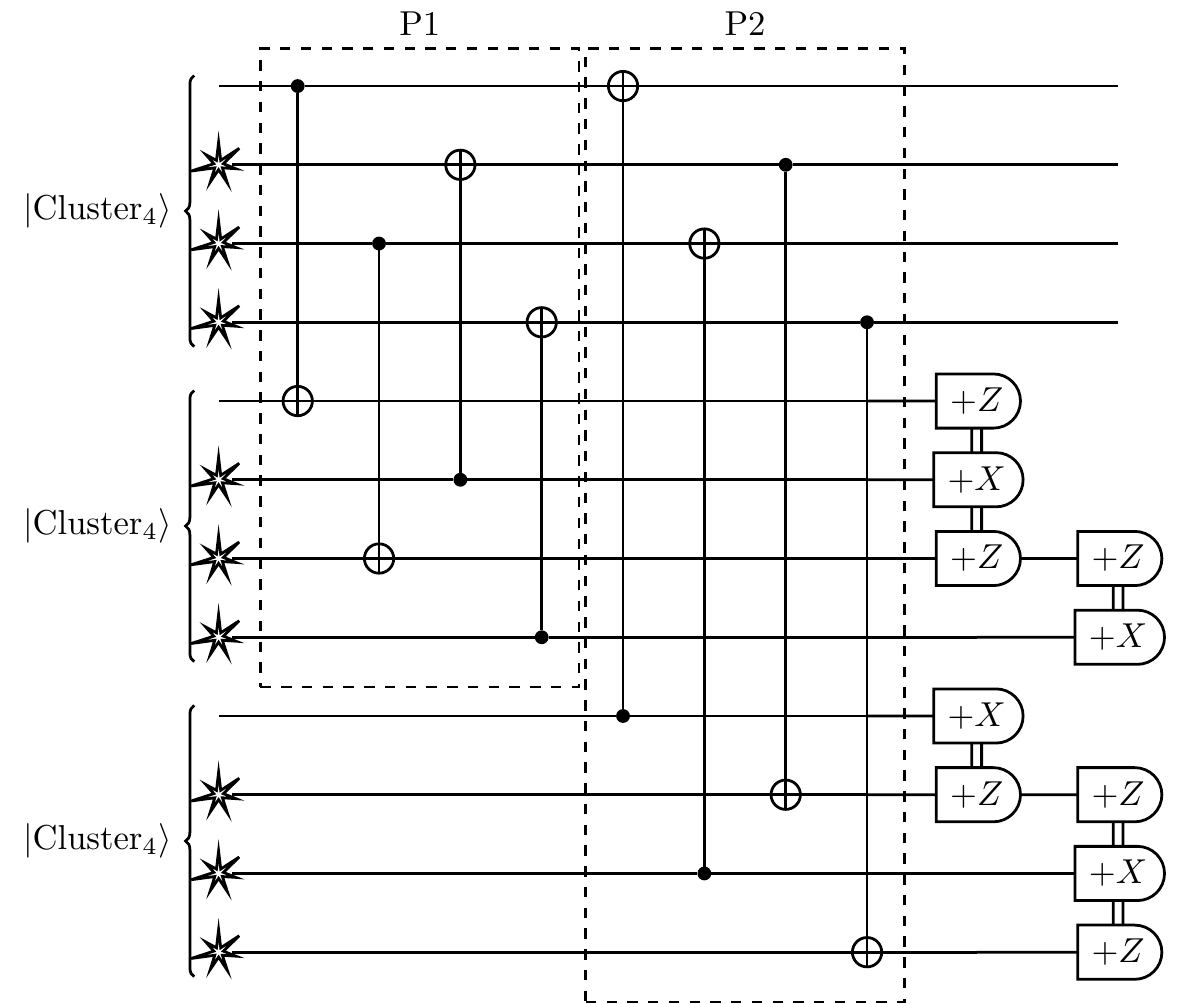}
    \includegraphics[scale=0.55]{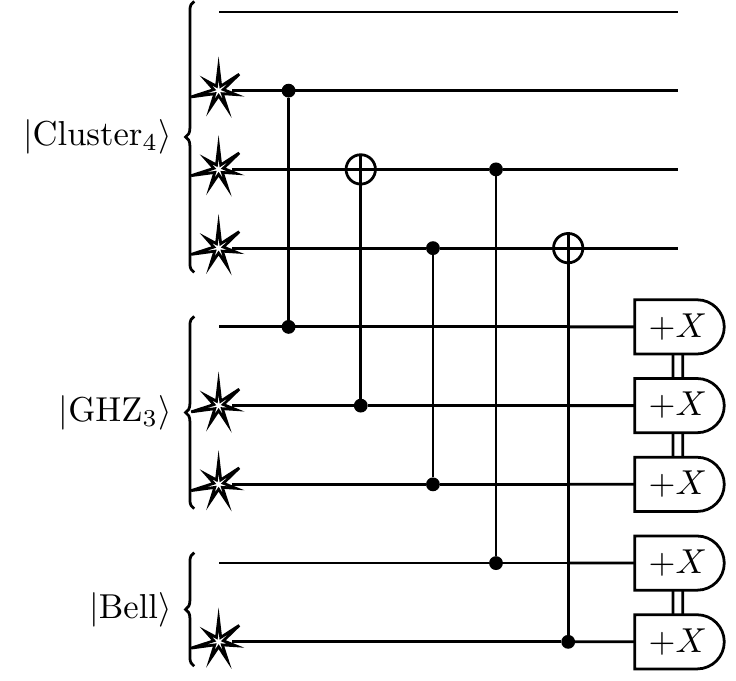}
    \caption{Purification circuits for the 4-qubit entangled states. Left-to-right, the circuits include our heterogeneous circuit for purification of a 4-qubit GHZ state; the standard two-stage circuit for the purification of the 4-qubit cluster state (we refer to each stage by the labels P1 and P2 -- out of these stages an asymptotic hashing purification protocol can be built), and our simpler heterogeneous circuit for purification of the 4-qubit cluster state. Our circuit provides higher fidelity of the final purified state and requires simpler sacrificial resources. Moreover, as our circuit is shorter, it is less susceptible to uncorrected gate errors. The vertical classical communication (double lines) between the measurement symbols denotes sets of classical bits whose parity needs to be checked in order to detect errors. Our circuits specifically target the measurement of stabilizers more probable to be flipped by network errors, given the particularities of the network (one qubit per state is never transmitted over the network and stays at the entanglement generation node).}
    \label{fig:4qs}
\end{figure*}

\section{Simulation Techniques}

A large class of quantum circuits, in particular circuits involved in quantum error detection and error correction, fall in the class of Clifford circuits which can efficiently be simulated on a classical computer~\citep{gottesman1998heisenberg}. The distinguishing feature of these circuits is that their application maps Pauli operators to Pauli operators, preserving the group structure. A natural way to track the evolution of a quantum state is to track how each of the basis vectors involved in the description of that state evolve. The number of basis vectors for $n$ qubits is $2^n$, which contributes to the difficulty of simulating generic quantum circuits on classical hardware. In the Heisenberg picture, this tracking corresponds to tracking the evolution of a complete set of commuting observables, e.g., a commuting subgroup of multiqubit Pauli operators, also of size $\mathcal{O}(2^n)$. Here the fact that Clifford circuits map Pauli operators into Pauli operators and preserve the group structure enable significant savings in the number of operators that need to be tracked. Instead of tracking an entire Pauli subgroup of size $\mathcal{O}(2^n)$, we can track a much smaller set of generators, i.e., operators that when multiplied can produce other elements of the subgroup. The size of the generating sets scales linearly with the number of qubits, enabling the efficient classical simulation of Clifford circuits.

In order to simulate noisy circuits we introduce a number of improvements to this method. A straightforward solution to simulating Pauli noise (including depolarization and biased noise) or erasure noise (for heralded or non-heralded erasures) is to perform a Monte Carlo simulation. In this approach the same circuit is simulated repeatedly, but at each gate and measurement there is a chance for an error to happen. The exact error and its probability are prescribed by the error model. The results of many such simulations are then averaged, giving an estimate of the performance of the noisy circuit.

A second approach we introduce is to perform a perturbative expansion in the error parameter as the circuit is being simulated. At each operation in the circuit we split and simulate both the "error branch" and "no error branch". We track the history of such splits and never take more than one turn in the "error branch" direction. We then average the results over all branches that have been taken. Thus, for a circuit of length $L$ we obtain a first-order perturbative expansion in the error parameter, while having to simulate only $\mathcal{O}(L)$ versions of the circuit for a high-precision estimate of the fidelity. At low error rates this is significantly faster than a Monte Carlo simulation of the same accuracy. Moreover, by using symbolic data structures, instead of floating point numbers, we can obtain analytic expressions for these results. Similarly to the Monte Carlo approach, any Pauli or erasure channel can be simulated in this manner. To our knowledge this is the first time such approach is suggested in the literature. The implementation of this algorithm is made available in the open source QuantumClifford.jl package.

\section{Small Circuits for Near-term Noisy Devices}

We will demonstrate purification circuits for a number of different states, including
\begin{align}
\ket{\mathrm{Bell}} & = \ket{00} + \ket{11} \\
\ket{\mathrm{GHZ}_3} & = \ket{000} + \ket{111} \\
\ket{\mathrm{GHZ}_4} & = \ket{0000} + \ket{1111},
\end{align}
as well as the four-qubit linear cluster state $\ket{\mathrm{Cluster}}$ which is defined by its stabilizer generators
\begin{equation}
\begin{split}
\hat{X}\hat{Z}\hat{I}\hat{I}\\
\hat{Z}\hat{X}\hat{Z}\hat{I}\\
\hat{I}\hat{Z}\hat{X}\hat{Z}\\
\hat{I}\hat{I}\hat{Z}\hat{X}.
\end{split}
\end{equation}

The states $\ket{\mathrm{GHZ}^\mathrm{c}_3}$ and $\ket{\mathrm{GHZ}^\mathrm{c}_4}$ are also introduced, which are the same as the aforementioned GHZ states, except for their first qubit being defined along the X axis instead of the Z axis. This change of reference frames is inconsequential, as a purification circuit for one of these states can be converted into a circuit for the other one by performing the reverse change of reference frames. Nonetheless, it is convenient to introduce this alternative reference frame, as that turns the GHZ states explicitly into bicolorable graph states, for which the canonical hashing method is defined~\citep{aschauer2005multiparticle}. We benchmark against that method.

\subsection{Heuristics for good small circuits}

The circuits we are investigating are of a small scale and exhaustive searches are not prohibitive to do on a classical computer. Nonetheless, there are a number of principles and heuristics that simplify the search. Previous works have shown the value in multiple selection circuits~\citep{fujii2009entanglement, nickerson2013topological, krastanov2019optimized}, where multiple sacrificial states are entangled together with the state to be purified before their measurement. Moreover, a crucial heuristic we exploited is the choice of stabilizer (of the state to be purified) that we desire to measure (by the sacrifice of another state). Picking stabilizers more sensitive to the error process (e.g., more probable to be triggered by an error) leads to a higher efficiency of error detection. This becomes particular important given the network error model we are studying: one of the qubits of each entangled state does not experience network errors because it is never transmitted over the network, therefore stabilizers involving that qubit are less probable to be triggered and are less valuable measurement targets. It is worth noting that the stabilizers of the sacrificial states needs to be consistent with the stabilizers of the state to be purified (similarly to the way graph states need to be similarly colored for a typical purification circuit~\citep{kruszynska2006entanglement}). This leads to the lack of sacrificial 3-qubit states in the purification of the 4-qubit GHZ state.

\subsection{Examples}

Fig.~\ref{fig:3qs} shows the typical 3-qubit GHZ circuit building blocks out of which the hashing purification protocol is built. It also shows a much simpler and smaller purification circuit which sacrifices simpler resources (2 Bell pairs instead of 2 GHZ states), performs at the same purified fidelity, and succeeds with much higher probability. The comparison of the two approaches, as well as subcircuits present in these two approaches can be seen in the first column of Fig.~\ref{fig:perf}. Crucial for the good performance of the simpler circuit is that distributing the raw Bell pairs over the network incurs less damage (only two of the qubits need to pass through the network, compared to four qubits that need to be networked when distributing two GHZ states). This approach is specifically optimized for small noisy quantum devices: once a larger number of GHZ states can, in principle, be sacrificed, then the hashing method will be able to overcome our advantage. But even in such a bright future, the lower level of the quantum network will benefit from using our cheaper optimized scheme as a pre-purifier for the GHZ states to be used by the higher layers of the purification stack. If the network error rate is $\varepsilon$  (the probability of $\hat{X}$, $\hat{Y}$, or $\hat{Z}$ error on a transmitted qubit is $\varepsilon$) then to first order the fidelity of a 3-qubit GHZ state before purification is $1-6\varepsilon$. Our circuit results in a success probability of $1-10\varepsilon$ and fidelity $1-2\varepsilon$. Compare this to a success probability of $1-16\varepsilon$ and the same fidelity in the case of the P1+P2 circuits used in the hashing method. By virtue of being smaller, our circuit has a higher probability of success while having the same final fidelity of purification. Moreover, the shortness of the circuit also makes it less susceptible to gate and measurement errors.

Similarly, Fig.~\ref{fig:4qs} showcases circuits for purification of various 4-qubit entangled states. Our 4-qubit GHZ purification circuit (evaluated in the second column of Fig.~\ref{fig:perf}) uses only 3 Bell pairs and performs at the same fidelity and higher success probability as the standard two-stage circuit that sacrifices two 4-qubit GHZ states. Moreover, our Cluster state purification circuit performs at even higher fidelities than the standard two-stage purification method, as seen in the last column of Fig.~\ref{fig:perf}, again with much humbler sacrificial resource requirements.

\subsection{Resilience to operational errors}

\begin{figure}
    \centering    \includegraphics[width=0.45\textwidth]{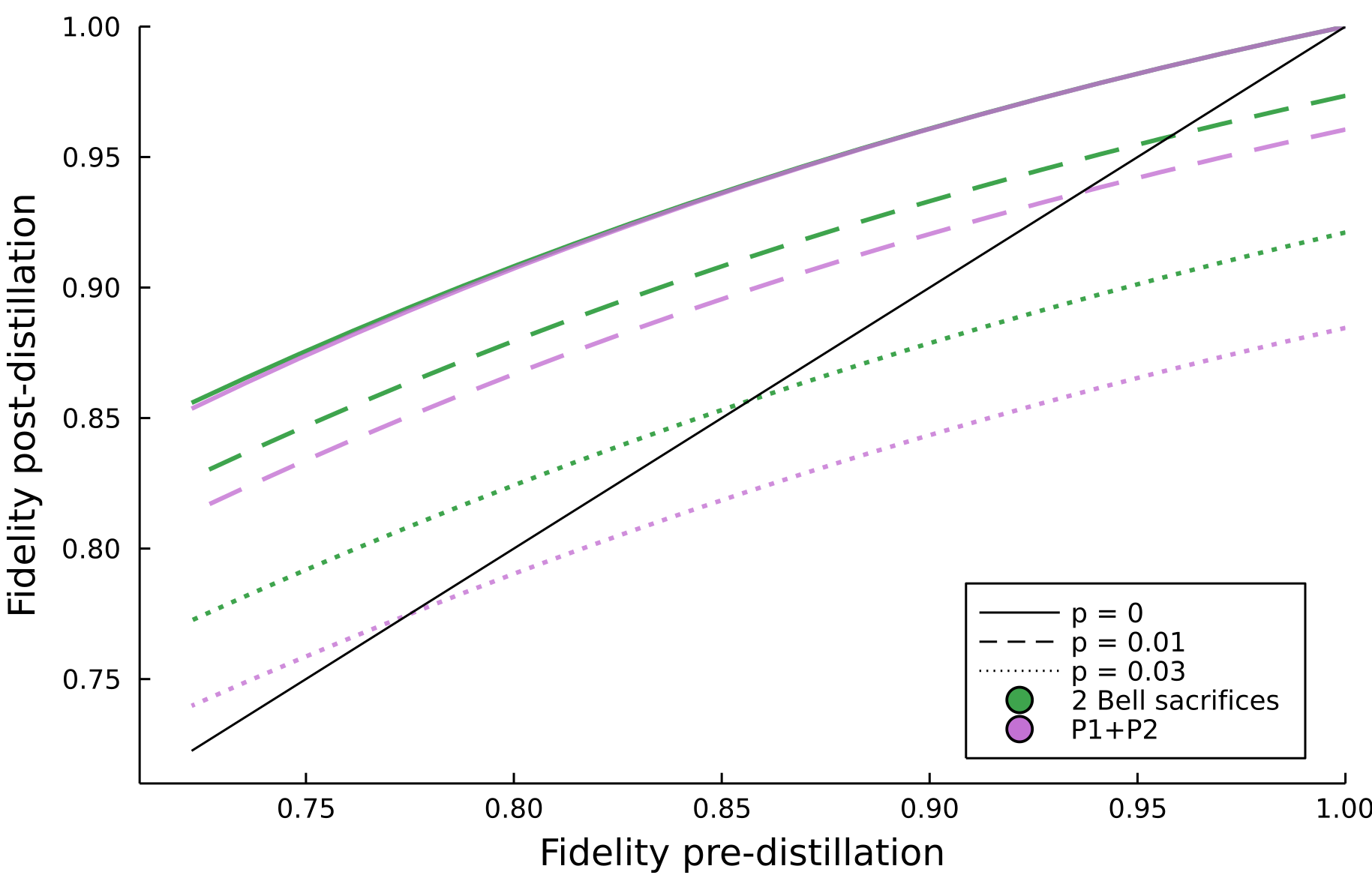}
    \caption{The effect of local gate infidelity is minimized in our purification circuits. Throughout this text we stressed the importance of designing small circuits for the sake of lowering resource overhead. This has the additional benefit of limiting the deleterious effects due to the finite fidelity with which local gates can be performed. This figure showcases the drastic improvement in purification fidelity in our circuits, simply by virtue of performing fewer imperfect operations. For this example we pick the circuits for purification of 3 qubit GHZ states. We compare a single round of the state-of-the-art hashing method (in purple) vs our smaller and simpler purification circuit (in green). With different line styles (solid, dashed, and dotted) we represent the infidelity (depolarization rate) of the local gates performing the purification.}
    \label{fig:gate_error}
\end{figure}

As alluded to above, the newly designed circuits presented here show additional advantages when imperfections in the local gates are considered. In Fig.~\ref{fig:gate_error} we benchmark one of our protocols in the presence of depolarization errors for each gate (at rate $p$) and measurement bit flip errors at the same rate. Our protocol, which uses fewer resources, performs at the same purification fidelity in the presence of perfect gates. However, due to its simplicity and shorter length it starts to significantly outperform the standard approach in the presence of imperfect gates.

\section{Conclusion}

There is a rich tradition in developing purification circuits for Bell pairs for small noisy quantum hardware~\citep{fujii2009entanglement,nickerson2013topological,nigmatullin2016minimally}, including even state-of-the-art optimization schemes providing circuits tailored to a given error model and hardware size~\citep{krastanov2019optimized}. Although elegant theoretical solutions like the hashing method exist, the aforementioned optimized small circuits are necessary when working on realistic noisy devices. The circuits presented here show this is still true in the case of multipartite entanglement purification. While obtaining the same or better purified fidelity, our circuits are shorter, sacrifice easier to obtain entanglement resources, and are less susceptible to gate errors. While we present examples of only GHZ and cluster state purification, this same approach can be used for the purification of arbitrary stabilizer states: locate the stabilizers most susceptible to being flipped by the network noise and employ simple purification circuits to check them. Running an optimizer~\citep{krastanov2019optimized} on top of this circuit generation would be particularly fruitful when targeting specific noise models on size-constrained hardware. This approach is, however, limited to small scale noisy circuits. Once fault-tolerant devices exist, many of the optimizations discussed here would be in principle unnecessary thanks to the availability of perfect gates. Nonetheless, the lower-level of such hardware implementation would still present a need for the small noisy circuits we have design, applied to the breeding of higher-quality entanglement to be used by the upper logical levels of the hardware.

\acknowledgments

Software created by the Julia open source community significantly aided this work and is gratefully acknowledged. We also thank Hyeongrak Choi, Dr. Michelle Victora, Prof. Don Towsley (University of Massachusetts), Prof. Saikat Guha (University of Arizona), and Dr. Michael Fanto (AFRL) for insightful discussions and comments on the work. S.K. is partially supported by the U.S. Department of Energy, Office of Science, Basic Energy Sciences (BES), Materials Sciences and Engineering Division under FWP ERKCK47. This work is also supported by the Harvard Physical Sciences Accelerator Award and Harvard Quantum Initiative Seed Grant. P.N. is a Moore Inventor Fellow and gratefully acknowledges support through Grant GBMF8048 from the Gordon and Betty Moore Foundation.

\bibliography{bib}

\end{document}